# Decentralized Adaptive Helper Selection in Multi-channel P2P Streaming Systems

SeyedAkbar Mostafavi and Mehdi Dehghan
Amirkabir University of Technology (Tehran Polytechnic)
IT & Computer Engineering Dept.
{a_mostafavi,dehghan}@aut.ac.ir

*Abstract.* In Peer-to-Peer (P2P) multi-channel live streaming, *helper* peers with surplus bandwidth resources act as micro-servers to compensate the server deficiencies in balancing the resources between different channel overlays. With deployment of helper level between server and peers, optimizing the user/helper topology becomes a challenging task since applying well-known reciprocity-based choking algorithms is impossible due to the one-directional nature of video streaming from helpers to users. Because of selfish behavior of peers and lack of central authority among them, selection of helpers requires coordination. In this paper, we design a distributed online helper selection mechanism which is adaptable to supply and demand pattern of various video channels. Our solution for strategic peers' exploitation from the shared resources of helpers is to guarantee the convergence to correlated equilibria (CE) among the helper selection strategies. Online convergence to the set of CE is achieved through the *regret-tracking* algorithm which tracks the equilibrium in the presence of stochastic dynamics of helpers' bandwidth. The resulting CE can help us select proper cooperation policies. Simulation results demonstrate that our algorithm achieves good convergence, load distribution on helpers and sustainable streaming rates for peers.

Keywords. Peer-to-Peer, Live Streaming, Helper Selection, Regret Tracking Algorithm, Correlated Equilibria

## I. INTRODUCTION

In the recent years, a large variety of multi-channel P2P streaming systems (like PPLive ; UUSee ; PPStream) with large users community have been successfully deployed which provide a wide range of high-quality video channels [1]. To guarantee high sustainable streaming rate for all users in all video channels in these systems is a challenging problem because of time-varying popularity of video channels and limited upload bandwidth of streaming servers. A well-known solution for this problem is utilizing the unused resources of so-called "*helper*" peers to alleviate the workload of streaming server and improve the quality of experience of end users [2][3]. On one hand, helpers act as micro-servers to compensate the deficiencies in upload bandwidth of server and decrease the resource-imbalance among the video channels. On the other hand, helpers as new service access points provide video streams to their connected peers.

With deployment of the intermediate helper level into the hierarchy of server and peers, new design questions arise that should be answered: "how should peers select the best helper considering the acceptable quality of service?" and "how should peers adaptively react to the changing available bandwidth of helpers?" The widely used solution for the peer selection problem in P2P systems is reciprocal peer choking (like tit-for-tat [4]) in which peers periodically replaces their worst neighbors with potential better ones. However, the reciprocity assumption is not valid in the helper selection problem because of one-directional stream delivery from helper to peers which makes the reciprocity-based approaches impracticable.

The problem of "helper selection" in a peer-assisted multi-channel P2P streaming system can be addressed by using either a centralized system-wide or distributed approach. The centralized approach is not applicable in the practical P2P streaming systems due to significant communication overhead and cooperative nature of solution in which peers should obey the decisions made by the central controller. On the other hand, distributed approach has low implementation complexity and low communication overhead and helper selection algorithms are implemented at the user side. It is also more adapted to the autonomous environments where peers make independent (and selfish) choice of the best helper node to connect to.

One of the driving features of peer-to-peer systems for adaptive learning algorithms is that these systems must operate in the changing conditions [5]. For example, utilities of peers in the P2P system may vary due to changes in streaming demands and helper upload bandwidth or join/leave of peers. It is therefore critical that peers detect and adapt to such events during their presence in the system. Given the non-stationarity induced by time varying upload bandwidth of helpers [6], the peers would incur connection costs in the form of degradation in the received streaming rate. The decentralized coordination of the helper selection would then require an adaptive procedure capable of both convergence to and tracking strategic equilibria as the environment changes.

To this end, we deploy the game-theoretic learning algorithm of *regret-tracking* [7][8], a stochastic-approximation-based re-expression of the well-known "regret-matching" algorithm [9][10], to guarantee the emergence and active tracking of correlated equilibria (CE) between the selection strategies of the peers. The game-theoretic foundation of our proposed scheme also relieves it from the node cooperation assumption which has been taken for granted in the comparable schemes. More specifically, we consider a multi-channel P2P streaming system scenario with several helper nodes among which users want to connect to. We show that to adhere a pure Nash equilibrium (NE) can cause serious stability problems in the system and therefore, we investigate a new concept, correlated equilibrium (CE), which is a more generic solution compared to the NE and usually leads to better performance in terms of system efficiency. We then propose a distributed algorithm based on regret-tracking for the users to adjust their strategies to converge to a set of correlated equilibria in a distributed manner.

Although concept of helpers has been addressed by a number of works [11][12][2][6][13][3], a fixed overlay topology was considered in most works and the user-helper topology was not optimized adaptively for improving system performance. The only related work is [14] which has proposed a solution for the problem of user-helper topology building. In this paper, authors propose a general optimization formulation for resource allocation in P2P streaming systems and then decompose the formulation to two separate optimization problem: storage/bandwidth allocation and topology-building optimization. A Markov approximation approach is applied to the overlay configuration problem and a distributed solution called "soft-worst-neighbor-choking" is proposed based on the optimal solution. The proposed algorithm cannot capture the inherent randomization in user join/leave behavior and stochastic dynamics of helpers' available bandwidth.

Our work differs with the existing work in that we develop an adaptive learning algorithm in which each peer adaptively deploy in a competitively optimal fashion given its limited information and orient the system converges to an equilibrium state in a distributed way. The proposed algorithm shows a good performance in terms of system efficiency and is especially adapted to the autonomous environments as P2P systems. We introduce cognition to the peers' choking decisions to enable proactive adaptation even when the game is of incomplete information and the environment dynamics (i.e. helper bandwidth fluctuations) is unknown; our solution works within the confines of bounded rationality, a practical assumption consistent with the limited capabilities of the peer nodes.

This paper is organized as follows. The theoretical background of work on game theory and no-regret learning are provided in the section 2. We then present our formulation for helper selection problem and discuss the details of regret-tracking algorithm for adaptive decentralized helper selection in Section 3. Section 4 is dedicated to the numerical analysis of proposed algorithm. We conclude the paper along with the presentation of some directions for future works.

## II. THEORETICAL BACKGROUND

Before presenting the proposed algorithm, we first provide a brief introduction on no-regret learning and regret tracking algorithm. Simply put, regret matching and regret tracking are procedures for learning correlated equilibrium strategies in a repeated game environment. Game players take actions, observe rewards, and adjust their strategies so as to guarantee convergence to a zero-regret condition, in which no player would adjust his past strategy based on present information.

*Regret tracking* is basically a stochastic approximation-based re-expression of the boundedly rational regret-matching algorithm, and is particularly suited for learning equilibria in slowly time-varying environments without Bayesian assumptions on the initial beliefs formed over the possible strategies of the opponents and states. While in the regret matching approach, the decisions of each player are based on the (uniform) average history of all past observed utilities, this choice is not desirable in our setting since the upload bandwidth state of helpers and thus the achieved streaming capacity evolve over time.

Regret-tracking algorithm is completely self-centered; players do not require any information about each others' actions or payoffs, and perform no explicit opponent modeling. Indeed, players need not even know of each others' existence. Therefore, these properties make it suitable for peer-to-peer applications. The cost of rendering the algorithm adaptive is correspondingly weaker convergence. Since convergence is to a correlated (as opposed to a Nash) equilibrium, it seems that there is a cooperative element to the algorithm, even though peers act solely on their own rewards.

The key to a regret-tracking-based strategy for a node is the way in which the past performance of its choices is assessed. Unlike regret matching, where all previous stages matter equally, here a node values more recent events higher than more distant events and regards the associated foregone utilities as less relevant. The ability to gradually let go of the past helps a node to better cope with time-varying environments. This is while incorporating all past events indiscriminately makes the strategy so rigid that eventually in response to changing conditions, a node would have no recourse but to forget all the past and start anew, with an empty history, rather than to maintain the original strategy. In order to track a non-stationary environment, regret tracking uses an exponentially, recency-weighted averaging scheme [15] which decreases the weight given to a perceived utility as the number of intervening utilities increases. The cost to be paid for such adaptability, however, is correspondingly weaker convergence in that the regret estimates never completely converge but continue to vary in response to the most recently experienced utilities. As a side note, it is worth mentioning that the aforementioned weighted averaging scheme can effectively be incorporated into regret-matching by compactly re-expressing the algorithm as a stochastic update equation with a constant step-size parameter [7][8], which will also allow for more efficient recursive implementation; here, we forfeit this compact re-expression and instead adhere to the original regret-matching-style description for the sake of clarity.

## III. PROBLEM FORMULATION: DECENTRALIZED HELPER SELECTION

We consider a helper selection scenario consisting of a P2P system with $H$ helper nodes and $N$ peers, in which each peer can choose the helper node to connect to. In such settings, the challenge for peers is to achieve maximum streaming capacity by selecting appropriate helper node. This scenario can be modeled as a non-cooperative game where players are the peers. Each player (peer) $i$ selects one helper among the available ones to maximize its utility function. We assume that the aggregate upload bandwidth of each helper is shared evenly among the peers connecting to it. In the following subsections, we present the helper selection game formulation and propose a distributed algorithm to achieve the equilibria of formulated game.

*A. System Model*

The repeated game for the helper selection problem can be described as follows.

**Players**. On the course of a system-wide helper selection process, every local ensemble of peers that currently aim to select a helper form a set of strategic players provided that their sets of contact points (helper sets) intersect. The symbol $N = \{1,2, \ldots, |N|\}$ denotes as representative set of such players. It worth mentioning that a peer in the helper selection game does not need to be explicitly aware of its fellow players; instead, it suffices to only infer its actually realized payoff at each stage.

**Actions**. The action space $A_i = \{h_1, h_2, \ldots, h_{|H|}\}$ for every peer $i$ consists of all the helper nodes are available to choose from. We denote by $x_i^n(a_i)$ the probability of node $i$ choosing $a_i$ at time $n$, and let $\chi_i := \Delta(A_i)$ be the set of probability distributions over $A_i$; i.e., $\chi_i := \{x_i | x_i(a_i) \epsilon [0,1], \sum_{a_i \epsilon A_i} x_i(a_i) = 1\}$. The mixed strategy of peer $i$ will then be: $x_i^n = [x_i^n(a_i)]_{a_i \epsilon A_i} \epsilon \chi_i$.

**Helper Selection Utility Function**. The instantaneous utility of a peer $i$ at stage $n$ of the helper selection game is a random variable. Each peer can directly calculate its received payoff based on the acquired streaming rate. Let $x \epsilon \mathbb{A}$ denote the joint helper selection decision. The peer $i$'s decision, $x^i$ is based on maximizing the expected value of a utility function $u^i(x)$ which reflects the upload bandwidth capacity of helper, the activity of other peers and the cost associated with connection to a given helper.

The key feature that characterizes our study as a *game* is that these considerations vary with the number of other active peers. We show these interactions as follows:

$$u_i^n = r_i^n = \frac{C_{h_j}^n}{\sigma_{h_j}^n}$$

where $r_i^n$ is the instantaneous streaming rate the peer $i$ receives from the connected helper, $C_{h_j}^n$ is the upload bandwidth capacity of helper $h_j$ at time step $n$, and $\sigma_{h_j}^n$ is the number of active peers which are connected to the helper $h_j$ at time step $n$.

**Histories**. A player's information consists of his past own-actions and perceived own-utilities. A private history of length $n$ for player $i$ is a collection $h_i^n = (a_i^0, u_i^0, a_i^1, u_i^1, \ldots, a_i^{n-1}, u_i^{n-1}) \epsilon H_i^n := (A_i \times \mathbb{R})^n$.

**Behavioral Strategies**. A behavioral strategy for player $i$ at stage $n$ of the helper selection game is a mapping: $\hat{\tau}_i^n: H_i^n \rightarrow \chi_i$. By considering $\hat{\tau}_i = (\hat{\tau}_i^n)_n$, the set of complete histories of the game after stage $n$ is: $H^n = (\prod_i A_i \times \mathbb{R}^{|N|})^n$.

**System-wide Objective**. Given a behavioral strategy profile $\hat{\tau}$, the utility of node in the infinite-horizon selection game is given by: $\lim_{N \to \infty} \sup \frac{1}{N} \sum_{n=0}^{N-1} \mathbb{E}_{\hat{\tau}}[u_i^n]$. Since we are interested in the expected helper selection game, we define its equilibria in terms of the strategy profile $x = (x_i^n)_n \epsilon \times_{i=1}^{|N|} \chi_i$. More specifically, $x$ is a mixed correlated equilibrium of the expected helper selection game if and only if:

$$\forall i \epsilon N, \sum_{a_{-i} \epsilon A_i} x(a_i, a_{-i}) \mathbb{E} u_i(a) \geq \\ \sum_{a_{-i} \epsilon A_i} x(a_i, a_{-i}) \mathbb{E} u_i(\hat{a}_i, a_{-i}), \forall \hat{a}_i \epsilon A_i \qquad (3-1)$$

where $a_{-i}$ denote $(a_j)_{j \epsilon N\{i\}}$.

*B. Helper Selection Through Regret Tracking*

In the non-cooperative helper selection game, utility function of peer $i$, $u_i$, after connection to helper $h_j$ is a non-increasing function of number of peers connecting to the helper $h_j$, which resembles the structure of utility functions in class of potential games [16]. The immediate result is that the helper selection game possesses a NE. But, playing the strategy prescribed by NE can yield poor performance in some situations. For example, consider $n$ peers that should choose among two helpers with the same upload bandwidth. Assume that in the first iteration, all peers are connected to the helper $h_1$. In the next iteration, utility of connection to the helper $h_2$ is higher than the helper $h_1$ because of its lower congestion. Therefore, all peers switch to the helper $h_2$. But this simultaneous switching makes the helper $h_2$ over-loaded and all peers will switch back to the helper $h_1$ in the next iteration. The switching back and forth between helpers will result in frequent interruption in the streaming flow and poor quality of experience.

Regret-tracking can be used to orient system to an equilibrium in decision-making scenarios under uncertainty. The uncertainty caused by stochastic dynamics is due to the evolution of some environmental state whose trajectory is not influenced by the players' decisions. In this paper, we are interested in the set of correlated equilibria of the helper selection game given that the players can indirectly acquire a coordination signal through the realized payoffs, and this coordination can lead to higher performance than if each player was required to act in isolation (as required by NE). Moreover, the convexity of the set of correlated equilibria arguably allows for better fairness between the peers, which is also evidenced by our simulation experiments. In the subsequent section, we rely on a reinforcement procedure to learn the expected payoffs simultaneously with the correlated equilibrium strategies.

Our adaptive heuristic, viz. "Regret-Tracking-based Helper Selection (RTHS)" learns the expected payoffs simultaneously with the CE strategies of the helper selection game. The symbols and definitions used in the algorithm are summarized in Table 1.

Using RTHS, the long-run helper selection game proceeds as follows:

- At stage 0, each peer $i \epsilon N$ has a random initial action $a_i^0$ and a zero *regret* value $Q_i^0$.

- At stage $n$, each peer $i \epsilon N$ chooses an action $a_i^n = a$ with a probability proportional to $Q_i^{n-1}(b,a)$; i.e., the regret for not having played $a$ instead of $b$. In other words, $Q_i^{n-1}(b,a)$ denotes the increase, if any, in the (weighted) average payoff that would result if all past plays of action $b$ were replaced by action $a$, and everything else remained unchanged. The peer $i$ then perceives a numerical value of its payoff $u_i^n$, that depends on the actions of other players.

- In the next stage, peer $i$ is required to update the estimate of its regret value $Q_i^{n-1}(a_i, a_{-i})$, $a_i \epsilon A_i$. The game moves on the stage $n+1$ and the process repeats.

Note that in the description of RTHS algorithm, the direct calculation of $\widehat{U}_n(x_{-i}^n)$ for other (unplayed) actions $x_{-i}^n$ may not be possible in general. In the followings, we present a variant of RTHS algorithm which estimate the quantity of $\widehat{U}_n(x_{-i}^n)$.

TABLE I. NOTATIONS

| Symbol | Definition |
|---|---|
| $n$ | time step |
| $N$ | number of peer nodes |
| $H$ | number of helper nodes |
| $p_i^n$ | peer $i$'s probabilistic mixed strategy at stage $n$ |
| $a_i$ | peer $i$'s selected action at stage $n$ |
| $u_i^n$ | instantaneous utility peer $i$ actually receives at stage $n$ |
| $\widehat{U}_n$ | estimated average utility for playing an alternate action |
| $\varepsilon$ | constant step size, $\varepsilon \epsilon [0,1]$ |
| $\mu$ | normalization constant |
| $Q_i^n(a,b)$ | peer $i$'s regret for not having played $a$ instead of $b$ |
| $Q_i^0$ | arbitrary initial regret |

Using RTHS, the nodes learn to play an equilibrium; i.e., after a given number of iterations, the strategy profile $x^n = (x_i^n, \ldots, x_{|N|}^n) \epsilon \Delta(A_1) \times \ldots \times \Delta(A_n)$ converges weakly to the set of correlated equilibrium behavior of the underlying game in a completely decentralized manner. Of particular note is that since in RTHS a peer does not need to perfectly monitor the others' actions, no particular synchronization mechanism is required between the participants. This relieves the algorithm from the exchange of signaling messages given that it only suffices to have an observation of the individual utilities per learning iteration.

---

Algorithm 1. Regret-Tracking-based Helper Selection (RTHS)

*Initialization:*
$Q_i^0 \coloneqq 0$; $a_i^0 \coloneqq$ Initial random action peer $i$ drawn from $\{h_1, h_2, \ldots, h_{|H|}\}$ with $p_i^0(a_i) = \frac{1}{|H|}$; $n \coloneqq 1$
**begin**
for $n = 1, 2, \ldots,$
  let $j = a_i^n$
  $u_i^n(j) = r_i^n$
  for each $k \neq j$
  $$Q_i^n(j,k) = \left[\widehat{U}_n(k) - \sum_{\tau \leq n} \varepsilon(1-\varepsilon)^{n-\tau} u_i^\tau(j)\right]^+$$
  $$p_i^{n+1}(k) = (1-\delta)\min\left\{\frac{1}{\mu}Q_i^n(j,k), \frac{1}{m^n - 1}\right\} + \frac{\delta}{m^n}, \quad k \neq j$$
  $$p_i^{n+1}(j) = 1 - \sum_{j \neq k} p_i^{n+1}(k)$$
  $n \coloneqq n+1$
**End**

---

The RTHS algorithm requires that both the selecting (unchoking) and not-selecting (choking) decisions made by peers be somehow evaluated at each stage of the game so as to be able to update the regret values associated with a peer's sequence of decisions. The unavailability of the information necessary for evaluating the alternate actions calls for a zero-knowledge learning scheme with bandit (or opaque) feedbacks. More specifically, a peer may define a *proxy* regret measure [20] by using the utilities it has perceived thus far when it actually played the alternate actions over the previous stages of the game. The calculation of the (*proxy*) regret measure $Q_i^n(j,k)$ would then require that the average $\widehat{U}_n(k)$ be estimated as follows:

$$\widehat{U}_n(k) = \sum_{\tau \leq n} \varepsilon(1-\varepsilon)^{n-\tau} \frac{p_i^\tau(j)}{p_i^\tau(k)} u_i^\tau(k) \qquad (3-2)$$

and then the average regret will be estimated as follows:

$$Q_i^n(j,k) = \left[\sum_{\tau \leq n} \varepsilon(1-\varepsilon)^{n-\tau} \frac{p_i^\tau(j)}{p_i^\tau(k)} u_i^\tau(k) - \sum_{\tau \leq n} \varepsilon(1-\varepsilon)^{n-\tau} u_i^\tau(j)\right]^+ \qquad (3-3)$$

In equation (3-2), $p_i^\tau$ denotes the play probabilities at stage $\tau$; in effect, the *proxy* regret for not having played $k$ instead of $j$ measures the difference of the average utility over the stages when $k$ was actually used and the stages when was $j$ used. The term $\frac{p_i^\tau(j)}{p_i^\tau(k)}$ normalizes the per-stage utilities so that the length of the respective stages would become comparable. Once again, our key modification to the estimation procedure discussed in [20] is to replace its simple (uniform) averaging basis by a weighted average to account for the time-varying randomness affecting the distribution of the perceived utilities.

Since it will consume too much resource to compute the estimated average regret directly according to equation (3-3), we design a recursive method to compute the estimated average regret. For each peer $i$, we first define a matrix $T_i^n$ with the following entries:

$$T_i^n(j,k) = \sum_{\tau \leq n; x_i^n = k} \frac{p_i^\tau(j)}{p_i^\tau(k)} u_i^\tau(x^n), \quad \forall j, k \epsilon A_i \qquad (3-4)$$

and then let the matrix evolve at the stage $n$ as follows:

$$T_i^n = T_i^{n-1} + \frac{u_i^n(x^n)}{p_i^n(x_i^n)} P_i^n \times e_{x_i^n} \qquad (3-5)$$

where $e_x = [0\ 0\ \ldots 1\ \ldots 0]$ with 1 in the $x^{th}$ position, and $P_i^n = [p_i^n(1)\ p_i^n(2)\ \ldots\ p_i^n(|H|)]^T$. Thus, the estimated average regret can be computed as follows:

$$Q_i^n(j,k) = \varepsilon * (T_i^n(j,k) - T_i^n(j,j))^+$$

Utilizing the equations (3-5) and (3-6), a recursive version of Algorithm 1 called "Recursive Regret-Tracking Helper Selection (R$^2$HS)" is presented in Algorithm 2.

IV. NUMERICAL RESULTS

In this section, we validate the feasibility and effectiveness of proposed algorithm by designing simulation scenarios in a P2P multi-channel streaming system. The dynamic helper selections strategies of each peer rely completely on the peer's local information, and therefore can be implemented in a fully distributed fashion.

**Algorithm 2. Recursive Regret-Tracking Helper Selection (R²HS) Algorithm**

*Initialization:*
$Q_i^0 \coloneqq 0$; $a_i^0 \coloneqq$ Initial random action peer $i$ drawn from $\{h_1, h_2, \dots, h_{|H|}\}$ with $p_i^0(a_i) = \frac{1}{|H|}$; $n \coloneqq 1$

**begin**
for $n = 1, 2, \dots,$
    for each peer $i$
        use play probabilities to select the action $j = a_i^n$
        compute the instantaneous utility $u_i^n(j) = r_i^n$
    for each peer $i$
        for each $k \neq j$
            use equation (3-5) to compute $T_i^n$
            use equation (3-6) to compute $Q_i^n$
        update play probabilities based as follows:
$$p_i^{n+1}(k) = (1-\delta)\min\left\{\frac{1}{\mu}Q_i^n(j,k), \frac{1}{m^n-1}\right\} + \frac{\delta}{m^n}, \quad k \neq j$$
$$p_i^{n+1}(j) = 1 - \sum_{j \neq k} p_i^{n+1}(k)$$
$n \coloneqq n + 1$
**End**

For reference, we formulate the helper selection problem as a cooperative optimization problem based on the Markov Decision Process (MDP) framework. Then, performance evaluation is done in terms of system social welfare, server and helpers' workload, individual peer's utility, and the balance in load distribution on helper nodes. The available bandwidth of helper nodes in each time step switches between three levels [700,800,900] according to a slowly changing random process.

### A. Alternative approach: Centralized MDP

When the cooperative optimization is considered over the set of centralized policies, then the problem is in fact of a single controller (the streaming server) which has all the information. In this setting, the cooperative policy $s(\mathbf{x}|\mathbf{y})$ is the probability that the server assigns the helpers $\mathbf{x} = (x_1, \dots, x_N)$ to the peers if the current helpers' states are given by the vector $\mathbf{y} = (y_1, \dots, y_N)$. We assume that the helper $i$ states can be modeled as an ergodic finite Markov chain $Y_i(t)$. The Markov chains $Y_i(t), i = 1, \dots, |H|$, are assumed to be independent. Let $\pi_i$ be the row vector of steady state probabilities of Markov chain $Y_i(t)$; let $\pi_i(y)$ be its entry corresponding to the state $y \in Y_i$. We also denote by the $\pi(\mathbf{x})$ the probability of state $\mathbf{x} = (x_1, \dots, x_{|H|})$. Since the Markov chains that describe the helper states are independent, $\pi(\mathbf{x}) = \prod_{i=1}^{|H|} \pi_i(x_i)$.

For a given $\mathbf{y} \in \mathbf{Y}$ and $\mathbf{x} \in \mathbf{X}$, the global occupation measure, $\rho^s(\mathbf{y}, \mathbf{x})$, is defined as

$$\rho^s(\mathbf{y}, \mathbf{x}) = \prod_{i=1}^{|H|} \pi_i(y_i) s(\mathbf{x}|\mathbf{y}).$$

For any given policy, $s$, and the corresponding occupation measure, $\rho^s(\mathbf{y}, \mathbf{x})$, we define the utility function of peer $i$ as follows:

$$R_i(s) = \sum_{\mathbf{y} \in \mathbf{Y}} \sum_{\mathbf{x} \in \mathbf{X}} u_i(\mathbf{y}, \mathbf{x}) \rho^s(\mathbf{y}, \mathbf{x})$$

where $u_i$ is the instantaneous utility of peer $i$. To find the social welfare of peers, the common objective of cooperative optimization for any policy $s$ is defined as

$$R(s) = \sum_{i=1}^{N} R_i(s)$$

$$u(\mathbf{y}, \mathbf{x}) = \sum_{i=1}^{N} u_i(\mathbf{y}, \mathbf{x})$$

The cooperative optimization problem is then formulated as follows:

$$\max_\rho R(s) \coloneqq \sum_{\mathbf{y} \in \mathbf{Y}} \sum_{\mathbf{x} \in \mathbf{X}} u(\mathbf{y}, \mathbf{x}) \rho^s(\mathbf{y}, \mathbf{x})$$

σ.τ. $\quad \sum_{\mathbf{x} \in \mathbf{X}} \rho(\mathbf{y}, \mathbf{x}) = \pi(\mathbf{y}) = \prod_{i=1}^{|H|} \pi_i(y_i)$

$$\sum_{\mathbf{y} \in \mathbf{Y}} \sum_{\mathbf{x} \in \mathbf{X}} \rho(\mathbf{y}, \mathbf{x}) = 1$$

$$\rho(\mathbf{y}, \mathbf{x}) \geq 0, \quad \forall \mathbf{x} \in \mathbf{X}, \forall \mathbf{y} \in \mathbf{Y}.$$

Figure 1 plots the evolution of regret value of worst player in a large-scale cooperative multi-channel P2P streaming. As seen, the regret value approaches to the zero, when the algorithm converges to the optimal solution.

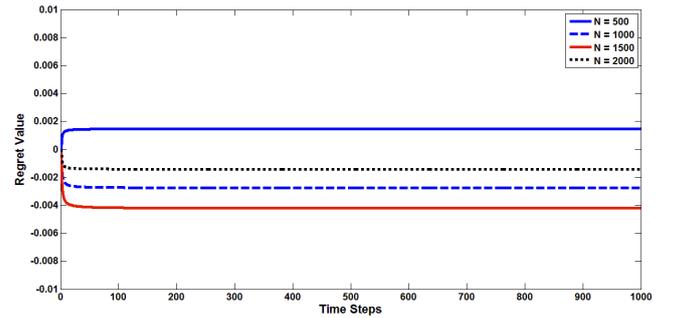

Fig. 1. Evolution of regret value of worst player in a large-scale scenario

Now consider a small-scale case with $N = 10$ peers and $|H| = 4$ helpers to evaluate the distributed RTHS algorithm against the centralized MDP algorithm as benchmark. As shown in the Figure 2, RTHS algorithm converges to the near-the-optimal solution for the dynamic helper selection game.

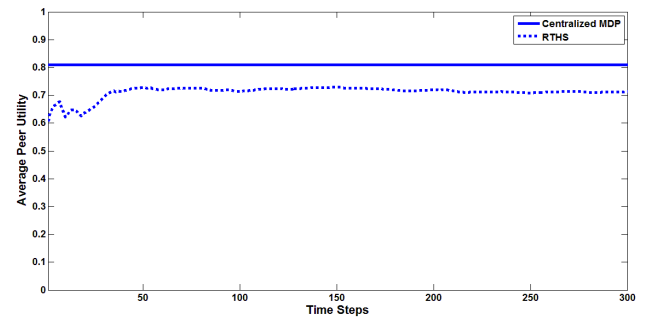

Fig. 2. The RTHS algorithm is a near-optimal algorithm in comparison with centralized MDP algorithm.

One of the features of RTHS algorithm is to evenly balance workload of peers on the helper nodes. This feature, as shown in the Figure 3, proves the fairness of RTHS algorithm which means the resources of helper nodes are fairly distributed among the competing peer nodes. So, all peers will receive a near equal share from pool of helpers' resources (Figure 4).

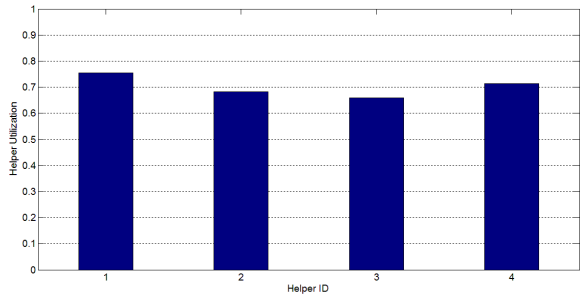

Fig. 3. The RTHS algorithm evenly distribute loads on the helpers.

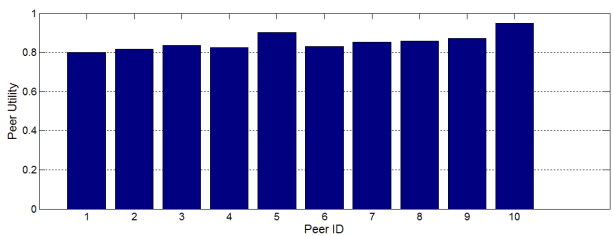

Fig. 4. The upload bandwidth of helpers is evenly distributed among peers.

Upload bandwidth provisioning through helpers decrease the workload on the streaming server. When the sum of peers' streaming demands exceeds from sum of helpers' provisioned bandwidth, the surplus requests are referred to the streaming server. The minimum bandwidth deficit of helpers is defined as the required amount of surplus bandwidth if the minimum upload bandwidth of all helpers is fully utilized. As shown in the Figure 5, the real server load is close to the minimum bandwidth deficit of helpers and therefore, helpers greatly decrease the load of streaming server.

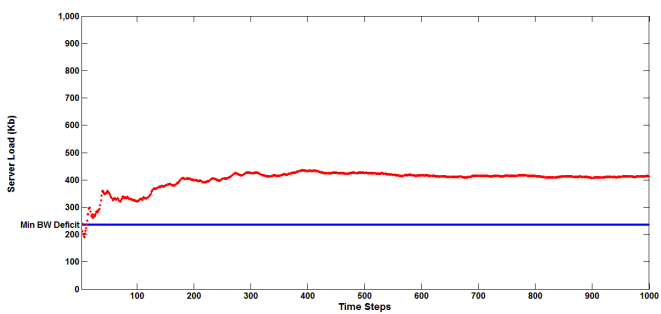

Fig. 5. The real server workload against the minimum bandwidth deficit of helpers.

## V. CONCLUSION

In P2P live streaming systems, the unused surplus upload bandwidth of peers can be utilized to alleviate the server load and increase the number of peers viewing the live channels. The helping peers called helpers play the role of micro-servers to the peers by sharing their upload bandwidth. The association of peers to the helpers become a challenging problem due to the unidirectional connection of peers to the helpers make the well-known reciprocity-based approaches (like tit-for-tat) unfeasible. In this paper, we formulate the helper selection problem as a non-cooperative game and deploy an online decentralized online learning algorithm called RTHS based on regret-tracking algorithm which converges to the set of CEs of helper selection game. The RTHS algorithm features in terms of quick convergence, even load distribution on the helpers and fair bandwidth allocation to the peers. It significantly reduces the server load and approaches to the optimal peer-helper association in comparison with the centralized MDP algorithm. The RTHS captures the stochastic dynamics in the helper upload bandwidth and adapts the selection decision of peers accordingly. In this work, we focused on a single channel system. Our future work is to extend the RTHS to the problem of joint bandwidth allocation in the helper level to the video channels and helper selection in the peer level.